\documentclass[aps,twocolumn,showpacs,preprintnumbers,nofootinbib,prd,superscriptaddress,groupedaddress]{revtex4}
\pdfoutput=1
\usepackage{amssymb,graphicx}
\usepackage{amsmath}
\usepackage{multirow}
\usepackage{epsfig}
\usepackage[usenames]{color} 
\usepackage[export]{adjustbox}
\usepackage{mathtools}

\newcommand{\beqn}{\begin{eqnarray}}
\newcommand{\eeqn}{\end{eqnarray}}
\newcommand{\beq}{\begin{equation}}
\newcommand{\eeq}{\end{equation}}

\def\mphi{m_{\phi}}

\def\rt{\tilde{\rho}}
\def\gammab{\hat{\gamma}}

\def\psib{\bar{\psi}}
\def\Phib{\overline{\Phi}}
\def\tg{\tilde{g}}

\def\tT{\tilde{T}}

\def\mcR{\mathcal{R}^2}
\def\meff{m_{\textrm{eff}}}

\begin{document}

\title{Generalized disformal coupling leads to spontaneous tensorization}
\author{Fethi M.\ Ramazano\u{g}lu, K{\i}van\c{c} \.I. \"Unl\"ut\"urk}
\affiliation{Department of Physics, Ko\c{c} University, \\
Rumelifeneri Yolu, 34450 Sariyer, Istanbul, Turkey }
\date{\today}

\begin{abstract}
We show that gravity theories involving disformally transformed  metrics in their matter
coupling lead to spontaneous growth of various fields in a similar fashion to the spontaneous
scalarization scenario in scalar-tensor theories. Scalar-dependent disformal transformations
have been investigated in this context, and our focus is understanding the
transformations that depend on more general fields. We show
that vector-dependent disformal couplings can be obtained in various different ways, each
leading to spontaneous vectorization as indicated by the instabilities in linearized equations
of motion. However, we also show that spontaneous growth is not
evident beyond vectors. For example, we could not identify a spontaneous growth mechanism 
for a spinor field through disformal transformations, even though there is a known example for
conformal transformations. This invites further work on the fundamental differences between
the two types of metric transformations. We argue that our results are relevant for
observations in strong gravity such as gravitational wave detections due to their promise of
large deviations from general relativity in this regime.
\end{abstract}
\maketitle

\section{Introduction}\label{intro}
Possible modifications to general relativity (GR) have been a topic of interest for
many decades, but until recently ideas in this line could only be tested in the weak-field
regime, where GR has been confirmed in all attempts~\cite{Will:2001LR}.
Gravitational wave (GW) observations are changing this picture, and dynamical strong-field
gravity can now be directly investigated~\cite{LIGOScientific:2018mvr,0264-9381-32-24-243001,Barack:2018yly}.
Despite these advances, the precision of GW detections is limited, which
has led to an increased interest in modifications of GR that provide large
deviations in strongly gravitating systems~\cite{PhysRevLett.116.221101}. 
Spontaneous scalarization in scalar-tensor theories where scalar fields grow near neutron stars
to provide nontrivial solutions provides exactly this type
of modification~\cite{PhysRevLett.70.2220}. This growth occurs due to a specific conformal
transformation of the metric in the matter coupling. In this study, we investigate
theories with disformal couplings that depend on fields beyond scalars (such as vectors),
and show that in many cases we can observe the spontaneous growth of the field.

Spontaneous scalarization contains a fundamental scalar degree of freedom
that governs gravity in addition to the metric tensor, that is, it is a scalar-tensor
theory. Any solution in GR is also a solution in the spontaneous scalarization
scenario, and it corresponds to a vanishing scalar field.
However, such solutions are unstable in the presence
of neutron stars~\cite{PhysRevLett.70.2220}. Arbitrarily small
scalar field perturbations go through exponential growth, and the eventual 
stable solution is a neutron star surrounded by a scalar cloud. The amplitude of the
scalar dies off away from the star, hence known weak field tests of 
gravity are satisfied. More strikingly, the value of the scalar field is large in
the vicinity of the neutron star, which leads to order-of-unity deviations from GR,
making spontaneous scalarization a prime target for strong gravity observations.

We will explain the basic mechanism of spontaneous scalarization and its generalizations 
in the following section, but the central idea is a tachyonic instability.
In the original theory of Damour and Esposito-Far\`{e}se (DEF) ~\cite{PhysRevLett.70.2220},
the matter fields couple to a metric that is conformally scaled by a function of the 
scalar (in the so-called Einstein frame). At the level of the scalar
equation of motion (EOM), this leads to an imaginary effective 
mass in the presence of matter. This is the famous tachyon, and it 
grows exponentially in time instead of oscillating.
The growth is quenched by nonlinear terms,
and the end point is a stable scalarized neutron star.

The essence of spontaneous scalarization is in an instability that is
eventually suppressed at large field values. It has been shown that a similar
mechanism exists in many other theories as well~\cite{Ramazanoglu:2017xbl,Ramazanoglu:2017yun}.
One idea to generalize spontaneous scalarization 
utilizes the fact that the scalar nature of the spontaneously 
growing field is not crucial. One can have a spontaneously growing vector
field as well, as long as there is a conformal scaling of the metric that 
is a function of the vector field, and the conformal function has a similar form to
that of the DEF theory~\cite{Ramazanoglu:2017xbl}. The spontaneous
growth idea applied to any field in this manner is named
``spontaneous tensorization''~\cite{Ramazanoglu:2017yun}.

Another place where spontaneous growth appears is a theory where
matter fields couple to a disformally transformed metric rather than a
conformally scaled one, whose technical details we will explain in the
next section.
Minamitsuji and Silva demonstrated that such a theory contains an
instability that causes spontaneous growth, and they also numerically
constructed explicit scalarized star solutions, but they did not 
consider spontaneous growth for other types of fields~\cite{Minamitsuji:2016hkk}. 

Our main task is combining the two aforementioned approaches that
generalize spontaneous scalarization.
Namely, we will study theories of gravity where disformal transformations play
a role, but these transformations are based on fields other than scalars. 
Such gravity theories have been in the literature as we will discuss
in more detail, but the fact that they give rise to spontaneous tensorization has
been overlooked to the best of our knowledge. Spontaneous growth generically leads to order-of-unity
deviations from GR, hence identification of its existence in any theory of 
gravity is especially important, since it dramatically increases the chances of
studying the theory using GWs or other means of strong field observations
that are becoming more commonly available. Our work
demonstrates that spontaneous growth is not merely a scenario specific to
a single theory, but it is a ubiquitous mechanism that exists in a wide variety of
gravity theories. On the other hand, we will show that not every theory of 
spontaneous growth with a conformal transformation can be automatically 
turned to one with a disformal transformation, and there are limits to known
mechanisms that provide spontaneous growth.

In Sec.~\ref{spontaneous_scalarization}, we give a basic 
explanation of spontaneous scalarization and its generalizations to both
disformally transformed metrics and non-scalar fields, basically summarizing the
literature. In Sec.~\ref{vector},
we present three different forms of disformal transformations that can lead
to spontaneous growth of vector fields. In Sec.~\ref{spinor}, we investigate
the spontaneous growth of spinor and rank-2 tensor fields through disformal couplings,
and we see that for various reasons results from vectors cannot be extended to all fields.
In Sec.~\ref{conclusion}, we summarize our results and their limitations,
discuss other related
theories such as disformal transformations beyond matter as in extended
Gauss-Bonnet gravity, and comment on connections to observations. We 
employ geometric units $G=c=1$ throughout the paper.

\section{Spontaneous scalarization through conformal and disformal couplings}\label{spontaneous_scalarization}
The first example of spontaneous growth in the gravity literature was
devised by DEF in scalar-tensor theories as in the action
\begin{align}\label{st_action}
  \frac{1}{16\pi} &\int d^4x \sqrt{-g}\ \bigg[R
 -\overbrace{ 2g^{\mu\nu}  \nabla_{\mu} \phi  \nabla_{\nu} \phi}^{T_\phi}\
 -\overbrace{2 m_{\phi}^2 \phi^2}^{V_\phi} \bigg] \nonumber \\
 &+ S_\text{m} \left[f_\text{m}, \tg_{\mu \nu} \right] \ ,
\end{align}
where 
\begin{align}\label{conformal_g}
\tg_{\mu\nu} = A^2(\phi) g_{\mu \nu} \ ,
\end{align}
and $f_\text{m}$ represents any matter degrees of freedom~\cite{PhysRevLett.70.2220}.
If the conformal coupling is of the form $A(\phi) = 1 +\beta \phi^2/2 + \ldots$,
such as the original choice $A(\phi)=e^{\beta \phi^2/2}$, 
$\phi=0$ is a solution that corresponds to GR,
but it is an unstable one in the presence of matter. When $\beta$
is negative and of the order of unity, neutron stars spontaneously grow 
scalar clouds around them that typically lead to large deviations
from GR. Scalarization weakens away from the star, guaranteeing
conformity with known tests of gravitation. Thus, investigation of such
modified theories is a realistic target for gravitational wave science.
We should add that the $V(\phi)$ term in Eq.~(\ref{st_action})
actually inhibits spontaneous growth, and it was not present in the original DEF theory,
but it is strongly favored to satisfy
recent binary star observations. This, and the details of other aspects of
spontaneous scalarization through this Lagrangian can be found
in Ref.~\cite{Ramazanoglu:2016kul}.

The origin of the instability can be seen in the linearized EOM for the scalar
\begin{align} \label{scalar_eom}
  \Box_g \phi &= \left( - 8 \pi A^4 \frac{d\left( \ln A(\phi) \right)}{d(\phi^2)} \tilde{T} + m^2_\phi \right)\phi \nonumber\\
  &\approx  ( - 4 \pi \beta \tilde{T} + m^2_\phi )\ \phi \equiv \meff^2\ \phi \ .
\end{align}
The trace of the stress-energy
tensor in the frame of $\tg_{\mu\nu}$ is negative as long as matter is
not ultrarelativistic, since $\tilde{T} = -\tilde{\rho}+3\tilde{p} \approx -\tilde{\rho}$.
So, for appropriate densities
and $\beta$ values, the effective mass $\meff$ is imaginary, which causes the
lowest-frequency Fourier modes of the scalar to have a tachyonic instability, since
$\omega^2 \sim \vec{k}^2+\meff^2 <0$. An equally
important fact is that this instability is eventually quenched by nonlinear effects
as the scalar grows to the nonperturbative regime, and we end up with a stable 
scalar cloud. 

To understand the first path to generalize the idea of DEF, note that
there is nothing specific to the scalar nature of the field in the mechanism
that incites the growth, or the nonlinear terms that later suppress it.
That is, one can replace the scalar, for example, with a vector $X_\mu$, as in the
action~\cite{Ramazanoglu:2017xbl}
\begin{align}\label{action_vt}
  \frac{1}{16\pi} &\int d^4x \sqrt{-g} \left[R - F^{\mu\nu} F_{\mu\nu} 
 -2m_X^2 X^\mu X_\mu \right] \nonumber \\
 &+ S_\text{m} \left[f_\text{m}, A_X^2(\eta) g_{\mu \nu} \right], \ \eta =g^{\mu\nu}X_\mu X_\nu \ ,
\end{align}
and still have spontaneous growth. Here, $F_{\mu\nu} = \nabla_\mu X_\nu -\nabla_\nu X_\mu$
and $A_X$ is an appropriate function of the vector field such as
$e^{\beta_X \eta/2}$. The vector EOM
\begin{align} \label{eom_vt_conformal}
\nabla_\rho F^{\rho \mu} = (-4\pi A_X^4 \beta_X \tilde{T} +m_X^2 )\ X^\mu 
\end{align}
has a tachyon-like nature as in the DEF theory, and observational signatures
of the theories are very similar. This line of thought can also be extended to
spinors~\cite{Ramazanoglu:2018hwk} and gauge bosons~\cite{Ramazanoglu:2018tig}. 

We should also add that the conformal coupling terms in these theories can depend
on the derivatives of the fields as well as the fields themselves. For example, in the
simplest case of the the scalar field, the action
\begin{align}\label{ghost_action}
\frac{1}{16\pi} \int &d^4x \sqrt{-g} [R-2\nabla_\mu\phi \nabla^\mu\phi -2\mphi^2 \phi^2]
\nonumber \\
+& S_\text{m}[f_{\textrm m},A_\partial^2(K) g_{\mu\nu}] \ , \ K=g^{\mu\nu} \partial_\mu \phi \partial_\nu \phi \ .
\end{align}
leads to the equation of motion~\cite{Ramazanoglu:2017yun}
\begin{align}\label{ghost_eom}
\nabla_\mu \left[(-8\pi\tilde{T}A_\partial^3 A_\partial'+1) \nabla^\mu \phi\right] = m_\phi^2 \phi \ .
\end{align}
This is radically different from the case in Eq.~(\ref{scalar_eom}), since there
is no modification to the mass term. However, for $A_\partial = e^{\beta_\partial K/2}$
with $\beta_\partial <0$ the principal part of the equation
(the part with the highest order of derivatives in the partial differential equation) reads
\begin{align}\label{ghost_eom}
(-4\pi\tilde{T}\beta_\partial+1) \Box \phi = \dots \ .
\end{align}
For large enough $\tilde{T}\beta_\partial$, it is the kinetic term rather than the 
mass-square term that changes sign, which means that we have a ghost-like
instability rather than a tachyonic one. This instability also grows exponentially from arbitrary
perturbations despite its different nature, which is the essence of spontaneous growth.
This is called ``ghost-based spontaneous scalarization.'' We can also
obtain ``ghost-based spontaneous vectorization'' by changing the dependence of
the conformal factor in Eq.~(\ref{action_vt}) $A_X \to A_F(F_{\mu\nu}F^{\mu\nu})$.

Another path to generalize spontaneous scalarization, and the one
we are going to examine in more detail in this study, uses
the fact that a conformal coupling is not the only way to obtain an
instability that causes spontaneous growth. The most general scalar-dependent disformal
transformation is~\cite{Bekenstein:1992pj}
\begin{align}\label{disformal_g}
\tg_{\mu\nu} = A^2(\phi) \left[g_{\mu \nu} +\Lambda B^2(\phi) \partial_\mu \phi \partial_\nu \phi \right] \ .
\end{align}
If we use this in the action Eq.~(\ref{st_action}), the resulting 
EOM is~\cite{Zumalacarregui:2012us,Minamitsuji:2016hkk}
\begin{align}
\label{scalar_eom_disformal}
\Box\phi
&= m_\phi^2 \phi+
\frac{4\pi}{1+\Lambda B^2 \partial_\mu \phi \partial^\mu \phi} \times \\
&\left\{
\Lambda B^2
\left[\left(\delta-\alpha \right) T^{\rho\sigma}_{}\partial_\rho \phi \partial_\sigma \phi
+
T^{\rho\sigma} \partial_\rho \partial_\sigma \phi \right] -\alpha T \right\} \ ,  \nonumber
\end{align}
where $\alpha(\phi) \equiv A^{-1}(dA/d\phi)$,
$\delta(\phi) \equiv B^{-1}(dB/d\phi)$, and the stress-energy tensor and its 
trace $T$ are in the frame of $g_{\mu\nu}$. The linearized EOM
arising from Eq.~(\ref{scalar_eom_disformal}) is more complicated than
the case of conformal coupling in Eq.~(\ref{scalar_eom}), but they were analyzed similarly
to the conformal case, which shows the existence of instabilities. 
Scalarized neutron star solutions for disformal couplings have
been explicitly constructed using numerical methods~\cite{Minamitsuji:2016hkk}. 

To understand the disformal transformation case better, first see that the
last $\alpha T$ term in Eq.~(\ref{scalar_eom_disformal})
arises from the overall conformal scaling $A^2$ in Eq.~(\ref{disformal_g}),
and behaves as an effective mass term as in tachyonic spontaneous scalarization.
The novel contribution of the disformal transformation can best be seen when 
we set $A(\phi)=B(\phi)=1$, in which case the principal part of the linearized EOM
becomes
\begin{align}\label{ghost_eom}
(-4\pi \Lambda T^{\rho\sigma} +g^{\rho\sigma})\ \partial_\rho \partial_\sigma \phi = \dots \ .
\end{align}
Hence, one can see that the character of the highest derivative term can change for
large enough $\Lambda$ and/or stress-energy density. This is very similar
to ghost-based spontaneous growth, but it arises from a completely different form
of coupling.

In the following sections, we are going to unite the two paths we discussed which generalize
spontaneous scalarization that is, using fields other than scalars, and using disformal rather
than conformal transformations. This way, we will investigate spontaneous tensorization of
vectors and other fields through disformal couplings that depend on these fields.

\section{Spontaneous growth from vector disformal couplings}\label{vector}
\subsection{Field disformal coupling}
The simplest disformal transformation of a metric by a vector field $X_\mu$ is
given by ~\cite{Kimura:2016rzw,Domenech:2018vqj}
\begin{align}\label{v_disformal}
\tg_{\mu\nu}= g_{\mu\nu} + B(x) X_\mu X_\nu
\end{align}
where $x=X_\mu X^\mu$. Here, we ignore the overall conformal scaling that is present in
Eq.~(\ref{disformal_g}) to concentrate our efforts on the purely disformal part of
the transformation.
We can devise a related modified gravity
theory in analogy to DEF given by the action
\begin{align}\label{vt_action}
\frac{1}{16\pi} \int &d^4x \sqrt{-g}\ [R-F_{\mu\nu}F^{\mu\nu}-2m_X^2 X_\mu X^\mu] \nonumber \\
+\int &d^4x \sqrt{-\tg}\ \mathcal{L}_\text{m}[f_\text{m},\tg_{\mu\nu}] 
\end{align}
where $F_{\mu\nu} = \nabla_\mu X_\nu -\nabla_\nu X_\mu$, and
we express the matter term explicitly in terms of the matter Lagrangian
$\mathcal{L}_\text{m}$. Replacing all occurrences of $\tg_{\mu\nu}$
with $g_{\mu\nu}$ corresponds to minimal matter coupling, hence GR.

Varying the action provides the EOM
\begin{align} \label{eom_vt}
\nabla_\mu F^{\mu\nu} = 
(m_X^2&-4\pi \sqrt{1+xB}\ B' \tilde{T}^{\rho\sigma}X_\rho X_\sigma) X^\nu \nonumber \\
&-4\pi \sqrt{1+xB}\ B \tilde{T}^{\mu\nu}X_\mu
\end{align}
where $B'=dB/dx$.
Here, the two stress-energy tensors defined with respect to the
bare and tilde metrics are related through
\begin{align} \label{Tin_frames}
T^{\mu\nu} &\equiv \frac{2}{\sqrt{-g}}\frac{\delta(\sqrt{-\tg} \mathcal{L}_\text{m})}{\delta g_{\mu\nu}} 
=\frac{2}{\sqrt{-g}}\frac{\delta \tg_{\rho\sigma}}{\delta g_{\mu\nu}}
\frac{\delta(\sqrt{-\tg} \mathcal{L}_\text{m})}{\delta \tg_{\rho\sigma}}  \nonumber \\
&= \sqrt{1+xB}\ ( \tilde{T}^{\mu\nu} - B' \tilde{T}^{\rho\sigma} X_\rho X_\sigma X^\mu X^\nu) \ .
\end{align}
Note that the lowering of the indices of the stress-energy tensors should be performed
with their respective metrics.

Our main interest is the spontaneous growth of $X_\mu$ in compact stars; hence;
we will have a closer look at the vector EOM in this setting. 
To the leading order in $X_\mu$, the EOM becomes
\begin{align} \label{eom_vt_linear}
\nabla_\mu F^{\mu\nu} \approx 
(m_X^2 \delta^\nu_{\ \mu} -4\pi B(0) \tilde{T}^{\nu\rho} g_{\rho\mu})X^\mu =\mathcal{M}^\nu_{\ \mu} X^\mu
\end{align}
where $\mathcal{M}$ can be interpreted as an effective mass-square tensor which is the
analog of $\meff$ in Eq.~(\ref{scalar_eom}). Then, all
it takes to have an instability is to have one negative eigenvalue of $\mathcal{M}$. For
matter that is not ultrarelativistic, the largest component of $\mathcal{M}$ is of 
magnitude of the rest mass density $\rt$ of the matter; hence, a negative mass mode
exists if $4\pi B(0) \rt \gtrsim m_X^2$, and suitable choices of $B$ lead
to spontaneous vectorization. 

Seeing the instability is easier for sufficiently symmetric
spacetimes where the metric and the stress-energy tensor are diagonal.
One common example is a spherically symmetric star with perfect fluid matter,
where the equation simplifies to
\begin{align} \label{eom_vt_linear2}
\nabla_\mu F^{\mu\nu} \approx 
(m_X^2 -4\pi B(0)\tilde{T}^{\nu\nu}g_{\nu\nu})X^\nu = \meff^2 X^\nu \ .
\end{align}
Here, the repeated indices are not summed on the right-hand side.
$\meff^2$ is clearly negative for appropriate choices of $B$,
and this indicates a tachyonic instability. 

Let us remember that an instability around the GR solution $X_\mu=0$
is desirable, but it is also essential that the instability shut off as it
grows so that the final solution is stable. Inspired by the DEF theory,
a natural choice is $B=\lambda_Xe^{\beta_X X_\mu X^\mu}$ for some constants
$\beta_X$ and $\lambda_X$. For example, in an astrophysical system where $X_0$ is
the dominant growing mode, $\beta_X>0$ ensures that the negative
contribution to $\mathcal{M}$ disappears as $X_0$ grows, killing the 
instability, while a $\lambda_X \sim 1$ would likely provide a powerful
enough instability in analogy to the DEF theory.

It should be clear that there is nothing magical about the exponential
form of $B$, and any function that behaves similarly when
$x=0$ and $x \to \infty$ provides spontaneous growth. 
However, ensuring that this recipe provides
stable neutron star solutions, that is, that the instability indeed shuts off eventually, 
requires more thorough numerical studies, such as
time evolution, which we will not attempt here.

It is curious to observe that the instability we have modifies the effective mass,
and is of tachyonic nature, unlike the scalar-dependent disformal coupling
in Eq.~(\ref{disformal_g}) which modifies the wave operator, leading to
a ghost-like instability. This difference is not due to the nature of the field,
but is related to the fact that the former directly uses the field in the disformal
transformation, whereas the latter necessarily uses the derivatives, since scalars
have no intrinsic indices. We will now see that ghost-like instabilities can arise for
vector-dependent disformal couplings as well, if the transformation includes the
derivatives of the field.

\subsection{Derivative disformal coupling}

It is also possible to have a derivative vector disformal coupling such as
\begin{align}\label{v_disformal_deriv}
\tg_{\mu\nu}= g_{\mu\nu} + \lambda B_F(x) F_{\mu\alpha} F_\nu^{\ \alpha}\ 
\end{align}
where $\lambda$ is a constant with dimensions of area that renders $B_F$ dimensionless.
This form of coupling has been discussed in the literature~\cite{Ezquiaga:2017ner},
but its consequences for any concrete theory, let alone in terms of spontaneous
growth, have not been investigated. We will assume $B_F(0)=1$ without loss of
generality.\footnote{This choice 
rules out $B(0)=0$, but we will soon see that this case is not relevant to the
discussion of spontaneous growth}

If we insert Eq.~(\ref{v_disformal_deriv}) into Eq.~(\ref{vt_action}), the
vector EOM in the resulting theory becomes
\begin{align} \label{eom_vt_deriv}
\nabla_\mu F^{\mu\nu} &= m_X^2 X^\nu \nonumber \\
&+4\pi \nabla_\mu [\sqrt{\chi} \lambda B_F (\tilde{T}^{\mu\beta} F_\beta^{\phantom{\beta} \nu} - \tilde{T}^{\nu\beta} F_\beta^{\phantom{\beta} \mu}) ]
 \nonumber \\
&-4\pi \sqrt{\chi} \lambda B_F' \tT^{\mu\beta} F_{\mu\alpha} F_\beta^{\phantom{\beta} \alpha}  X^\nu \ .
\end{align}
In this and all the following cases with disformal coupling, we define the ratio of the 
determinants of the metrics in the two frames as
 \begin{align}
 \sqrt{\chi} \equiv \sqrt{-\tilde{g}}/\sqrt{-g}\ .
 \end {align}

The nature of this equation is less transparent compared to
Eq.~(\ref{eom_vt}), but we can have a better idea by first linearizing, and
then concentrating on the principal part, i.e. considering only the highest 
derivative terms
\begin{align} \label{eom_vt_deriv2}
[-4\pi \lambda (\tT^{\rho\sigma}g^{\mu\nu}- \tT^{\nu\sigma}
g^{\rho\mu}) +g^{\rho\sigma}g^{\mu\nu}]\ \nabla_\rho F_{\sigma\mu} =\dots .
\end{align}
Note that the terms arising from the disformal transformation, $\tT^{\rho\sigma}g^{\mu\nu}- \tT^{\nu\sigma}
g^{\rho\mu}$, generically would not vanish, and hence would change the overall
sign of the kinetic term $\nabla_\rho F_{\sigma\mu}$ for appropriate (large enough)
choices of $\lambda$. The "wrong" sign kinetic term simply means that there is a ghost-like
instability in such regions of spacetime similar to the case of scalar-dependent disformal
transformation in Eq.~(\ref{disformal_g}). This is not a surprise since we know that ghost-like
instabilities arise from derivative couplings, which is the case in both theories.

The ghost-like instability can be more easily seen in
specific cases such as when
the metric and the stress-energy tensor are diagonal as for a
nonrotating neutron star with perfect fluid matter. Let us also assume
that the rest mass density, and hence $\tT^{00}$,
is dominant for ease of analysis. Consider the $\partial_t^2$ terms in the EOM
for the spatial components $\nu=i$ 
\begin{align} \label{eom_vt_deriv3}
[-4\pi \lambda B_F(0) \tT^{00}+g^{00}]g^{ii}\ \partial_t^2 X_i  \approx \dots \ \ \textrm{(no sum)}
\end{align}
where the right hand side contains at most first time derivatives. The coefficient of
$\partial_t^2 X_i$ can reverse its sign in the presence of matter. For this to happen,
$\lambda \gtrsim \rt^{-1}$ should be satisfied. Our assumptions might look too restrictive
since the pressure terms can be comparable to density terms in $\tT^{\mu\nu}$,
especially for more massive neutron stars. This would not qualitatively change our conclusions 
since $\tT^{\rho\sigma}g^{\mu\nu}- \tT^{\nu\sigma} g^{\rho\mu}$ does not vanish
in general, and has a value of $\sim\rt$, hence $\lambda \gtrsim \rt^{-1}$ would
still be sufficient with order of unity changes of $\lambda$.

Just as in Eq.~(\ref{v_disformal}), we still need the instability to shut off
as the field grows. This can again be satisfied by nonlinear terms in $X_\mu$,
namely, a decaying function $B_F=e^{\beta_F x}$ with a choice of
sign for $\beta_F$ that ensures that $B_F$ vanishes for growing values
of $X_\mu$, or $B_F=e^{\bar{\beta}_F x^2}$ for $\bar{\beta}_F<0$ would ensure
that the initial instability around $X_\mu=0$ would vanish for larger fields.

The form of the disformal term in Eq.~(\ref{v_disformal_deriv}) is inspired by the
standard kinetic term $F_{\mu\nu}F^{\mu\nu}$, in a similar fashion to the
relationship between the standard Proca potential term $X_\mu X^\mu$
and Eq.~(\ref{v_disformal}). Other choices such as 
replacing $F_{\mu\nu}$ with nonsymmetric $\nabla_{\mu} X_{\nu}$
might seem possible, but they may lead to unregularized ghosts in flat
space limit, hence we choose to avoid them~\cite{deRham:2014zqa}.
Such terms are in general possible with carefully chosen couplings
in the most general vector-tensor theories which contain at most two derivatives
to avoid Ostrogradsky's theorem~\cite{Heisenberg:2014rta,Kimura:2016rzw}.
Our choice in Eq.~(\ref{v_disformal_deriv}) is to demonstrate the relevance of
derivative vector disformal couplings, especially in the context of spontaneous
tensorization, and we will not attempt to construct the most general disformal
vector coupling in this study. 

Lastly, we remind that $m_X$ does not play an essential role in our discussion,
which means the case $m_X=0$ still possesses the instability. 
Such a theory preserves the gauge symmetry $X_\mu \to X_\mu +\partial_\mu \rho$
for any scalar $\rho$, which might be desirable depending on the physical
interpretation of $X_\mu$.

\subsection{Disformal coupling through the Abelian Higgs mechanism}
So far we have considered the intrinsically massive vector field, the Proca field,
in Eq.~(\ref{vt_action}).
A second, and physically better motivated way of introducing mass
to a vector field is the Abelian Higgs mechanism which preserves the
gauge symmetry of the massless vector.
This mechanism is given by the following action
\begin{align}\label{higgs_action}
\frac{1}{16\pi} \int &d^4x \sqrt{-g}\ [R-F_{\mu\nu}F^{\mu\nu}] \nonumber \\
-\frac{1}{16\pi}\int &d^4x \sqrt{-g}\ \big(2 \overline{D_\mu \Phi} D^\mu \Phi +2V(\Phib\Phi)\big) \nonumber \\
+\int &d^4x \sqrt{-\tg}\ \mathcal{L}_\text{m}[f_\text{m},\tg_{\mu\nu}] 
\end{align}
where $\Phi$ is a complex scalar, $D_\mu \Phi = (\nabla_\mu - i e X_\mu) \Phi$
is the gauge covariant derivative with a coupling constant $e$,
and an overbar means complex conjugation.
The gauge transformation is $X_\mu \to X_\mu -\nabla_\mu \rho$ and 
$\Phi \to e^{ie\rho}\Phi$. The Higgs mechanism introduces the vector
field mass through the hidden $e^2\overline{\Phi}\Phi X_\mu X^\nu$ in the
scalar kinetic term. This is thanks to the choice
\begin{align}
V(\Phib\Phi) = m_0^2(u^2 - \Phib\Phi)^2/(2u^2)\ ,
\end{align}
which causes the ground state of $\Phi$ to attain a nonzero value.

The trivial matter coupling choice that preserves the gauge symmetry is
\begin{align}\label{higgs_disformal_deriv}
\tg_{\mu\nu}= g_{\mu\nu} + \lambda_D B_D(\overline{\Phi}\Phi)  \overline{D_{(\mu} \Phi} D_{\nu)} \Phi\ ,
\end{align}
where $_{(\ )}$ represents symmetrization, and we choose the normalization $B_D(0)=1$.
This disformal transformation contains both scalar and vector dependences through the gauge
covariant derivative $D$.
Similar disformal transformations have been investigated in the cosmology
literature~\cite{Papadopoulos:2017xxx}, but its effects in terms of spontaneous tensorization
of compact objects is a novel concept to the best of our knowledge.
The action in Eq.~(\ref{higgs_action}) together with a conformal, rather than disformal,
transformation for $\tg_{\mu\nu}$ is known to cause spontaneous growth of vector and
gauge boson fields~\cite{Ramazanoglu:2018tig}.

The EOMs for the scalar and vector fields arising from Eq.~(\ref{higgs_disformal_deriv}) are
\begin{align}\label{eom_higgs}
&\nabla^\nu F_{\nu\mu} =\Delta^\nu_{\ \mu}
(e^2 \Phib\Phi\ X_\nu + J^\Phi_\nu) \ , \nonumber \\
&\Theta^{\mu\nu}
[\nabla_\mu\nabla_\nu-e^2X_\mu X_\nu -2ieX_{\mu} \nabla_{\nu}-ie\nabla_{\mu} X_{\nu} \big] 
 \Phi \nonumber\\
 &-4\pi \nabla_\mu (\lambda_D B_D T^{\mu\nu}) D_\nu \Phi \nonumber \\
 &=[m_0^2(\Phib\Phi/u^2-1)+\mu_\Phi^2]\Phi 
\end{align}
where $J^\Phi_{\mu} =ie (\Phib\nabla_\mu \Phi - \Phi \nabla_\mu \Phib )/2$ and
\begin{align}\label{higgs_aux}
\Delta^\nu_{\ \mu} &=-4\pi\lambda_D B_D T^{\nu}_{\phantom{\nu}\mu} +\delta^\nu_{\ \mu} \nonumber \\
\Theta^{\mu\nu} &= -4\pi\lambda_D B_D T^{\mu\nu}+g^{\mu\nu} \\
\mu_\Phi^2& = -4\pi\lambda_D B_D' T^{\mu\nu} \overline{D_{\mu} \Phi} D_{\nu}\Phi \ . \nonumber
\end{align}

Eq.~(\ref{eom_higgs}) behaves qualitatively similarly to the spontaneous growth
cases we have seen so far. $\Delta^\nu_{\ \mu}$ and $\Theta^{\mu\nu}$ cause the
principal parts of the equations for $X_\mu$ and $\Phi$ to change sign when
$\lambda_D$ is large enough, leading to ghost-like instabilities in both. Note that
$\Phi$ also gets a contribution to its effective mass through $\mu_\Phi^2$ that
can potentially drive a tachyonic instability for an appropriate form of $B_D$,
but this term only appears beyond the linear order in perturbations of $\Phi$
around its equilibrium value $\Phib\Phi=u^2$. This means it does not initiate
spontaneous growth, but it can play a role once the fields grow to a level where
nonlinear effects are dominant. 

Remember that we require the shutoff of the instability as the fields grow, which suggests that we need
$B_D$ to decay as $\Phi$ grows. Inspired by our experience, $B_D=e^{\beta_D \Phib \Phi}$
with $\beta_D<0$ is a possible choice. Even though we considered a $B_D$ that is only a function of $\Phi$, it can
be generalized to include vector dependence, $B_D(\Phi, X_\mu X^\mu)$,
which would bring new effective mass terms to the vector as well. However,
these would not change the qualitative picture of the spontaneous tensorization process we
described.

The case for the spontaneous growth of a non-Abelian gauge field $W^a_\mu$
is very similar to its Abelian version. The action is given by
\begin{align}\label{action_ym_higgs}
\frac{1}{16\pi} \int &d^4x \sqrt{-g} [R-F^{a\mu\nu} F^a_{\mu\nu}] \nonumber\\
-\frac{1}{16\pi}\int &d^4x \sqrt{-g}  \big[ 2(D_{\mu} \Phi)^\dagger D^\mu \Phi +2V(\Phi^\dagger\Phi) \big]
 \nonumber \\
+\int &d^4x \sqrt{-\tg}\ \mathcal{L}_\text{m}[f_\text{m},\tg_{\mu\nu}] 
\end{align}
where $^\dagger$ indicates the Hermitian conjugate, and
\begin{align}
F^a_{\mu\nu} &= \nabla_\mu W^a_\nu - \nabla_\nu W^a_\mu + e f^{abc} W^b_\mu W^c_\nu \nonumber \\
V(\Phi^\dagger \Phi) &= \frac{1}{2}\frac{m_0^2}{u^2}(u^2 - \Phi^\dagger\Phi)^2 \ .
\end{align}
The Higgs field $\Phi$ is now a multidimensional object  that can be acted upon by $T^a$, 
generators of the Lie algebra of the gauge group. $a,b,c$ label $T^a$, and the structure constants
$f^{abc}$ are defined as  $[T^a,T^b]=if^{abc}T^c$. Then, the disformal transformation
\begin{align}\label{higgs_disformal_deriv_ym}
\tg_{\mu\nu}= g_{\mu\nu} + \lambda_W B_W(\Phi^\dagger \Phi)  D_{(\mu} \Phi^\dagger D_{\nu)} \Phi\ ,
\end{align}
results in a theory where the non-Abelian fields grow spontaneously.

\section{Disformal coupling beyond vectors}\label{spinor}
We have seen that spontaneous growth arising from disformal couplings
can be easily adapted to vectors. However, the idea is even more general,
and we can consider disformal coupling of any field. We will investigate the
cases of spin-half and spin-$2$ particles in this section, and see that our approach
to scalars and vectors does not proceed as smoothly in all cases in terms
of obtaining spontaneous growth phenomena.
\subsection{Spinor disformal coupling}
Our extension of spontaneous growth through disformal coupling from scalars
to vectors can be generalized to other fields. As in the vector case, we can get
inspiration from spontaneous growth through conformal transformation, where
the next targets after vectors were spinors~\cite{Ramazanoglu:2018hwk}. Spontaneous
growth of spinor fields in gravity is less known, hence we will try to summarize all the
basic aspects of a spinor-dependent conformal metric scaling first, which will be crucial in 
understanding the spinor-dependent disformal coupling and its role in spontaneous growth.

Consider the following action
\begin{align}\label{spinor_action}
\frac{1}{16\pi} \int &d^4x \sqrt{-g}\ R \nonumber \\
+\frac{1}{16\pi}\int &d^4x \sqrt{-g}\ \left[ \left(\psib \gamma^\mu (\nabla_\mu \psi) 
- (\nabla_\mu \psib) \gamma^\mu \psi \right)
-2m \psib \psi \right] \nonumber \\
+\int &d^4x \sqrt{-\tg}\ \mathcal{L}_\text{m}[f_\text{m},\tg_{\mu\nu}] 
\end{align}
where $\psi$ is a Dirac bispinor and $\psib \equiv -i\psi^\dagger \gammab^0$
is constructed with the flat space gamma matrix $\gammab^0$. 
Definitions for the flat space gamma matrices $\gammab^{(\mu}\gammab^{\nu)}=\eta^{\mu\nu}$,
curved space gamma matrices $\gamma^{(\mu}\gamma^{\nu)}=g^{\mu\nu}$,
covariant derivatives $\nabla_\mu$ for spinors and other
relevant mathematical details can be found in Ref.~\cite{Ramazanoglu:2018hwk}.
The second line is simply the action for a minimally coupled spinor field in gravity.
The spinor field spontaneously grows in the presence of matter if we have a conformal
coupling of the form $\tg_{\mu\nu} = A_\psi^2 g_{\mu\nu}$ in the matter action with
\begin{align}\label{Apsi0}
A_\psi = e^{\beta_\psi (\psib \gammab^5 \gamma^\mu (\nabla_\mu \psi) 
- (\nabla_\mu \psib) \gammab^5 \gamma^\mu \psi )/4}
\equiv e^{\beta_\psi \mathcal{L}_\psi^{5,K}/2}
\end{align}
for a constant $\beta_\psi$ and
\begin{align}\label{gamma5_curved}
\gamma^5 \equiv 
\frac{i}{4!} \epsilon_{\mu\nu\rho\sigma} \gamma^\mu\gamma^\nu\gamma^\rho\gamma^\sigma
=\frac{i}{4!} \tilde{\epsilon}_{abcd} \gammab^a\gammab^b\gammab^c\gammab^d = \gammab^5\ ,
\end{align} 
which gives the EOM
\begin{align}\label{dirac_eqn_spinorization}
\gamma^\mu \nabla_\mu \psi -  
\frac{\mathbb{I} - \zeta_{\psi} \gammab^5}{1-\zeta_{\psi}^2}\ 
[m-(\nabla_\mu \zeta_\psi) \gammab^5\gamma^\mu/2]\ \psi &= 0 \ ,
\end{align}
with $\zeta_{\psi} \equiv 4\pi \tilde{T}\beta_\psi A_\psi^4$.

To understand why the above EOM leads to spontaneous 
growth, let us examine the purely tachyonic spinor EOM
in flat space~\cite{Chodos:1984cy,Jentschura:2011ga}
\begin{align}\label{dirac_eqn_tachyon}
\left( \gammab^\mu \partial_\mu - \gammab^5m\right) \psi =0 \ .
\end{align}
Let us investigate a plane wave solution 
$\psi = u(\vec{k}) e^{ik_\mu x^\mu}=e^{-i\omega t+i\vec{k}\cdot \vec{x}}$
of this equation
\begin{align}\label{tachyon_dispersion}
&(\gammab^\nu \partial_\nu - \gammab^5 m)(\gammab^\mu \partial_\mu -\gammab^5 m) \psi = 0\nonumber \\
\Rightarrow &\left[-\eta^{\mu\nu} k_\mu k_\nu 
-i (\gammab^\mu\gammab^5+\gammab^5\gammab^\mu) m k_\mu 
+\gammab^5 \gammab^5 m^2\right] \psi =0 \nonumber \\
\Rightarrow &\ \omega^2 =\vec{k} \cdot \vec{k} -m^2 \ ,
\end{align}
where we used $\gammab^\mu\gammab^5+\gammab^5\gammab^\mu=0$. This
clearly chows that $\omega$ is imaginary for large wavelength (small $|k|$)
modes, and leads to exponential growth rather than oscillation in time. This
is the instability mechanism of the tachyon, hence the name ``tachyonic
Dirac equation''. 

We have a mix of tachyonic and non-tachyonic 
terms in Eq.~(\ref{dirac_eqn_spinorization}), but this equation also has the exponential
growth modes for large enough $\lambda_\psi$, details of which
can be found in Ref.~\cite{Ramazanoglu:2018hwk}.

One important
aspect of Eq.~(\ref{dirac_eqn_spinorization}) is that it contains
the term $\nabla_\mu \zeta_\psi$. $\zeta_\psi$ itself contains
derivatives of $\Psi$  (since it is a function of $A_\psi$
in Eq.~(\ref{Apsi0}) ), which means $\nabla_\mu \zeta_\psi$
has second derivatives of $\psi$, seemingly  becoming the
principal part  of the EOM. This would be a radical change since 
the principal part has a dominant effect on the behavior of the equation, as
we have utilized so far. However, using the EOM to express the
derivative terms in Eq.~(\ref{Apsi0}) leads to the expression
\begin{align}\label{Apsi}
A_\psi^4 = \frac{\zeta_{\psi}}{4\pi \beta_\psi \tilde{T}}=\exp \left(-2 m \beta_\psi \psib \psi
\frac{ \zeta_{\psi}}{1-\zeta_{\psi}^2} \right)\ ,
\end{align}
that is, $\zeta_\psi$ can be written as a function of $\psi$, albeit implicitly, rather than
its derivatives. This means the equation of motion is still a first
order partial differential equation.\footnote{The  $\nabla_\mu \zeta_\psi$
term was missing in the original publication, and is added in a recent erratum.}
We will look at this important fact once more for the disformal coupling case.

The task at hand is finding a disformally transformed $\tg_{\mu\nu}$ that
can possibly lead to spontaneous spinorization. Examination of the relationship between
conformal and disformal transformations that lead to spontaneous growth for
scalars and vectors immediately suggests that we lower one of the contracted 
indices in Eq.~(\ref{Apsi0}) and add such a term to the metric
\begin{align}\label{spinor_disformal}
\tg_{\mu\nu}= g_{\mu\nu} + \lambda_\psi B_\psi \left[ 
\psib \gammab^5 \gamma_{(\mu} \nabla_{\nu)} \psi 
- (\nabla_{(\mu} \psib) \gammab^5 \gamma_{\nu)} \psi \right],
\end{align}
where $B_\psi(\psib\psi)$ can be normalized as $B_\psi(0)=1$.
Varying the action gives
\begin{align}\label{eom_spinor}
(\zeta^{\mu\nu} \gamma^5 + g^{\mu\nu}\mathbb{I}) \gamma_{\mu} \nabla_{\nu} \psi &- 
(\mu_\psi-(\nabla_\mu \zeta^{\mu\nu})\gamma_\nu) \psi =0
\end{align}
where
\begin{align}\label{spinor_aux}
\zeta^{\mu\nu} &= 8\pi\sqrt{\chi}\lambda_\psi B_\psi \tT^{\mu\nu} \\
\mu_\psi& =m -4\pi\sqrt{\chi}\lambda_\psi B_\psi' \tT^{\mu\nu}\left[ 
\psib \gammab^5 \gamma_{\mu} \nabla_{\nu} \psi 
- \nabla_{\mu} \psib \gammab^5 \gamma_{\nu} \psi \right]\ . \nonumber
\end{align}

At first sight, for large enough $\lambda_\psi$, the $\gammab^5$ term may seem to 
dominate over $\mathbb{I}$ in Eq.~(\ref{eom_spinor}). An instability
occurs around $\psi=0$ in a similar manner to our tachyonic dispersion
relation in Eq.~(\ref{tachyon_dispersion}). 

The above explanation of the tachyonic nature of Eq.~(\ref{eom_spinor}) 
overlooks an important fact: the equation of motion contains the 
$\nabla_\mu \zeta^{\mu\nu}$ term. Note that  $\zeta^{\mu\nu}$ already
contains derivatives of $\psi$ through $\chi$. For conformal transformations,
we were able to use the equation of motion to express $\zeta_\psi$ without
any derivatives, as in Eq.~(\ref{Apsi}), hence $\nabla_\mu \zeta_\psi$ stayed a first order term in 
the differential equation. 

We were not able to perform a similar procedure for $\zeta^{\mu\nu}$,
that is, to the best of our knowledge, $\zeta^{\mu\nu}$ depends on the derivatives of
$\psi$, which means $\nabla_\mu \zeta^{\mu\nu}$ contains second derivatives
of $\psi$. In other words, $\nabla_\mu \zeta^{\mu\nu}$ is the principal part of the 
EOM, Eq.~(\ref{eom_spinor}). This means that even if its coefficient is small, it
has a leading role in the time evolution of $\psi$, and as a consequence the above 
analysis for the existence of a tachyonic degree of freedom cannot be 
repeated verbatim. A change in the order of the EOM as a partial differential
equation (from first to second in this case) is radical, and has never been the
case in any of the spontaneous growth theories so far.

Moving beyond the above major modification to the spinor EOM, 
we were also not able to find a clear tachyonic mode in this second order differential 
equation. This is an important difference from the case of vectors where the instability
directly appears both in conformal and disformal transformations. 

We should add that any spontaneously growing spinor 
should be considered as a classical object as opposed to a quantum field~\cite{Ramazanoglu:2018hwk}. 
This is mainly because spinor fields obey the Pauli exclusion principle when quantized,
which means that their occupation numbers cannot have arbitrary values.
Spontaneous spinorization would result in a continuously adjustable spinor field value
which is determined by the spinorizing object (e.g. the neutron star), 
hence the exclusion principle cannot be accommodated. A purely classical spinor is not
a commonly encountered object, and its spontaneous growth also calls for care
in understanding the basics of half-integer spins in nonquantum contexts~\cite{Endlich:2013spa}.

\subsection{Spin-2 disformal coupling}
Can we have a disformal transformation based on a spin-2 field, since it seems 
to be the natural choice after scalars (spin-0) and vectors (spin-1)? The answer 
is negative based on our current knowledge of interacting spin-2 field
theories.

To start with, we should note that there is no known theory of spontaneous growth
of a spin-2 field, even when we have a conformal transformation~\cite{Ramazanoglu:2017yun}. 
A good starting point to understand this is looking at the known interacting spin-2 theories, 
which have only recently been developed~\cite{deRham:2014zqa}.
Since the metric is also a spin-2 field, such theories can be considered as theories of
two metrics $g_{\mu\nu}$ and $f_{\mu\nu}$ (bimetric theories), whose generic action is given by 
\begin{align} \label{action_bigravity}
S = S_{EH}(g) + S_{EH}(f) + S_{int}(f, g) 
+ S_m \left[f_m, g_{\mu \nu} \right]
\end{align}
where $S_{EH}$ is simply the Einstein-Hilbert action for a given metric,
and the interaction term between the metrics $S_{int}$ has to be of a 
specific form in order to avoid undesirable ghosts~\cite{PhysRevLett.106.231101,Hassan2012}.
We also included a minimal matter coupling $S_m$ in one of the metrics, which
is known to preserve the ghost-free nature of the theory.

Looking at our previous examples of spontaneous growth through nonminimal
matter coupling, the most straightforward attempt to induce an instability 
in this case would be replacing the metric in the matter coupling with a transformed 
one
 \begin{align} \label{matter_transformation_spin_2}
S_m \left[\psi_m, g_{\mu \nu} \right] \to S_m \left[\psi_m, \tg_{\mu \nu} \right] \ ,
\end{align}
where $\tg_{\mu\nu}$ is a function of both $g_{\mu\nu}$ and $f_{\mu\nu}$.
For example, a possible disformal transformation is
 \begin{align} \label{disformal_spin_2}
\tg_{\mu\nu} = g_{\mu\nu} + \lambda_f f_{\mu\nu}\ .
\end{align}
That is, matter field couples to both metrics at the same time. However, coupling
to a single metric is crucial for having a ghost-free theory, and the ghost is known
to reappear with composite metric couplings as in Eq.~(\ref{disformal_spin_2})~\cite{deRham:2014fha}. 
Thus, stable spontaneous growth that comes with such matter couplings seems
to be impossible. One potential way out of this can be a scenario where the ghost 
only comes in at extremely high energy scales and the theory can be reconciled with
observation~\cite{deRham:2014fha}. Whether this happens for our
proposed disformal transformation is not clear. We note that
conformal transformations also contain couplings to both metrics in a similar fashion, and
seem to fail due to the same reason~\cite{Ramazanoglu:2017yun}. Overall, all known 
coupling forms that lead to spontaneous growth seem to fail for spin-2 fields.

In summary, it is not straightforward to generalize disformal transformation-based
spontaneous growth beyond vectors. This was not a surprise for spin-2 fields, since spontaneous
growth is not known for any form of coupling. However, the spinor case is puzzling, since
spontaneous spinorization does occur for conformal transformations. Understanding the deeper reasons
for such differences between conformal and disformal transformations is an important part of future
studies on spontaneous growth in gravity.

\section{Conclusion}\label{conclusion}
The original spontaneous scalarization theory of DEF is the quintessential example of
spontaneous growth in gravity where large deviations from GR in strong fields
provide an ideal target for GWs. This theory has a scalar-dependent conformal
transformation in the metric that couples to matter, which provides an imaginary
effective mass, hence the growth. This idea was recently generalized in various ways,
and here we investigated the interaction of two such paths. First, one can replace the scalar in the 
conformal scaling of DEF with other fields, e.g. a vector, and obtain
spontaneous tensorization for general fields. Second,
one can replace the scalar-dependent conformal scaling of the metric with a still
scalar-dependent disformal transformation, and obtain a novel form of spontaneous
scalarization.

In this study, we combined the two approaches above, and showed that spontaneous
growth also occurs when the scalar dependence of the disformal transformation is
generalized to other fields. We have devised three vector-based
disformal transformation theories, and showed that they generically possess the instabilities 
that incite spontaneous growth. The first of these theories can be said to be somehow simpler
than the scalar-dependent case since the transformation only contains the field, and not its
derivatives. This is also reflected in the fact that the instability in this theory is tachyon-like,
while all other cases contain a ghost-like instability which can lead to astrophysically unusual
structures~\cite{Ramazanoglu:2017yun}.

We have also showed that ideas to generalize spontaneous growth have their limitations.
It was already known that the usual conformal transformations cannot be used to
obtain a theory of spontaneously growing spin-2 field, and this continues to be the case
for disformal transformations due to similar reasons. The case is more curious for 
spinor fields. Even though one can obtain spontaneous spinorization based on
conformal transformations, it is not clear whether this happens for their disformal counterparts.
Disformal coupling changes the nature of the equation of motion for the
spinor from a first order partial differential equation to a second order one. This radical
change also makes it hard to establish that there is an instability in the linearized equations,
though we have not ruled out this possibility either.

Before we conclude, let us discuss one of the paths to generalize the spontaneous scalarization of
DEF we ignored so far: using couplings beyond the matter term. It has been recently
shown that a scalar-dependent coupling to any term in the Lagrangian, for
example the Gauss-Bonnet terms as in the action
\begin{align}\label{egb_action}
\frac{1}{16\pi} \int d^4x \sqrt{-g} [R-2\nabla_\mu\phi \nabla^\mu\phi
+\lambda^2 f(\phi) \mcR]\ ,
\end{align}
leads to the spontaneous growth of the scalar~\cite{Silva:2017uqg,Doneva:2017bvd,Antoniou:2017acq}. 
Here $\mcR = R^2-4R_{\mu\nu}R^{\mu\nu}+R_{\mu\nu\rho\sigma}R^{\mu\nu\rho\sigma}$
is the Gauss-Bonnet invariant of $g_{\mu\nu}$, and since it is nonzero purely due to curvature,
one can obtain spontaneous growth near black holes as well as neutron stars.
$\mcR$ can be replaced with other curvature- or field-dependent  terms~\cite{Herdeiro:2018wub},
or the scalar can be replaced with another field such as a vector~\cite{Ramazanoglu:2019gbz},
and one can still obtain more general spontaneous tensorization phenomena. 
Following the theme of the current study, we can hope to find spontaneous growth
in the analogues of these theories with disformal transformations, e.g.
\begin{align}\label{egb_action_disformal}
\frac{1}{16\pi} \int d^4x \sqrt{-g} [R-2\nabla_\mu\phi \nabla^\mu\phi
+\lambda^2 \tilde{\mathcal{R}}^2]\ ,
\end{align}
where $\tilde{\mathcal{R}}^2$ is the Gauss-Bonnet invariant of a disformally
transformed metric $\tg_{\mu\nu}$. Unlike the above case, the coupling of the 
scalar (or vector) field to the Gauss-Bonnet term is not straightforward anymore.
Even though such theories have been considered in the past~\cite{Zumalacarregui:2013pma},
their equations of motion are quite complicated, and we were not able to
find clear signs of spontaneous growth with our linearized analysis. More thorough 
studies may shed more light on this issue.

We have taken the path of considering a single form of coupling and investigating 
its dependence on different fields, but there are alternative directions to explore the
landscape for theories that feature spontaneous growth phenomena.
For example, one can consider
the most general gravity theory that contains a given field in addition to the metric,
and then consider all possible coupling terms for the field. Such theories for scalar fields 
were pioneered by Horndeski~\cite{Horndeski1974},
and all possible mechanisms of spontaneous scalarization in this case
have been recently investigated~\cite{Andreou:2019ikc}.
This procedure can be repeated for other fields, which would form
a systematic approach that would complement ours.

This study can be considered as a demonstration of the fact that
spontaneous growth is widespread in gravity theories. Theories that have very
different action formulations can have very similar behavior if they contain similar
dynamical mechanisms. In our case,
the mechanism is spontaneous growth based on an instability, which is eventually
regularized due to nonlinear interactions. The similarity is not merely a theoretical one,
observational signatures of these theories are also quite alike, and particularly prominent 
in the context of GWs. Thus, we think it will be fruitful to consider spontaneous
tensorization theories as members of the same family as far as modifications
to GR are considered, despite the diversity in their Lagrangian formulations.
On the other hand, there are limitations to which fields can grow spontaneously by
a given coupling in the action,
and discovering the underlying reasons for these will be an important aspect of
understanding spontaneous growth in gravity.

\acknowledgments
We thank Masato Minamitsuji and Hector O. Silva for their help on 
scalar-based disformal couplings. The authors are supported by
Grant No. 117F295 of the Scientific and Technological
Research Council of Turkey (T\"{U}B\.{I}TAK).
We also acknowledge networking support by the GWverse COST
Action CA16104, ``Black holes, gravitational waves and fundamental physics.''

\bibliography{/Users/fethimubin/research/papers/references_all}

\end{document}